\documentclass{revtex4-1}

\usepackage{graphics}

\begin{document}

\newcommand{\bmb}[1]{\mathbf{#1}}
\newcommand{\U}{{\sf U}}
\renewcommand{\S}{{\sf S}}
\newcommand{\bbmb}[1]{\mathbf{#1}}
\newcommand{\ms}{m^{*}}
\renewcommand{\S}{{\sf S}}
\renewcommand{\dag}{+}
\newcommand{\bbm}{\bbmb}

\title[]{Plasmons in finite spherical ionic systems}

\author{W. Jacak}

\affiliation{Institute of Physics, Wroc{\l}aw University of Technology,
Wyb. Wyspia{\'n}skiego 27, 50-370 Wroc{\l}aw, Poland,\\ 
 \email{witold.aleksander.jacak@pwr.edu.pl}  }

\date{Received: date / Accepted: date}

\begin{abstract}
The challenging question on possible plasmon type excitations in finite ionic systems is discussed. Related theoretical model is formulated and developed in order to describe surface and volume plasmons of ion liquid in finite electrolyte systems. Irradiation of ionic surface plasmon fluctuations is studied in terms of the Lorentz friction of oscillating charges. Attenuation of surface plasmons in the ionic sphere is calculated and minimized with respect to the sphere size. Various regimes of approximation for description of size effect for damping of ionic plasmons are determined and a cross-over in damping size-dependence is demonstrated. The most convenient, optimal dimension of finite electrolyte system for energy and information transfer by usage of ionic dipole plasmons is determined. The overall shift of size effect to micrometer scale for ions in comparison to nanometer scale for electrons in metals is found, as well as the red shift by several orders of plasmonic resonances in ion systems predicted  in a wide range of variation depending on ion system parameters. This convenient opportunity of tuning resonances differs properties of ionic plasmons from plasmons in metals where electron concentration is firmly fixed.
\end{abstract}


\maketitle

\section{Introduction}
\label{intro}
Recent experimental and theoretical investigation of plasmon oscillations in metallic nanoparticles focused attention on the fundamental character of this phenomenon and also on great prospects of applications. In particular, the so-called plasmon effect in solar cells modified in nano-scale with on surface deposited metallic particles leads to the significant growth of their efficiency \cite{wzr3a,wzmocn1,wzr2,konk,wzmocn2,mof}. The mediating role in collecting of sun-light energy is played by surface plasmon oscillations in metallic nanoparticles due to their radiative properties. Irradiation of energy of plasmon oscillations is preferable for energy transport applications. As it was observed experimentally and later predicted theoretically, irradiation losses of plasmon energy are strongly sensitive to the metallic nanoparticles size \cite{jacak11}.
  
Strong irradiation of plasmon oscillations in metallic nanoparticles plays also a major role in construction of plasmonic wave-guides with high transference efficiency. Many experimental studies \cite{atwater1,vidal2010} indicated that  periodic linear structures of metallic nanoparticles serve as efficient plasmon wave-guides with low damping \cite{zastos,maradudin,deabajo}. The wave-lenghts of propagating in such structures plasmon-polaritons typically are by one order shorter in comparison to light with the same frequency, which allows for avoiding diffraction limits in optical circuits \cite{plasmons,pitarke2006,berini2009}. This is perceived as a way to forthcoming constructions of plasmon opto-electronic nano-devices, not attainable by using only light wave-guides limited by diffraction constraints. An efficient energy transfer in plasmon wave-guides is also conditioned by radiative properties of surface plasmons in metallic nano-components. 

Radiative losses of  plasmon oscillations can be described by the so-called Lorentz friction \cite{lan,jackson1998}. Accelerating charges irradiate electro-magnetic  wave and the related energy loss can be accounted for  as an effective electric field which hampers electron movement. For the case of  the oscillating dipole as for the  dipole-type surface plasmons in a metallic nanosphere, the Lorentz friction force is proportional to the third order time-derivative of this dipole \cite{lan}. 
Let us emphasize here that the strong irradiation of surface plasmons in metallic nanospheres, linked to the Lorentz friction, is exclusively present in sufficiently large metallic nano-particles. Small  metallic nano-paricles  in form of the clusters of size $1-5$ nm do not exhibit irradiation efficiency so high as nanospheres with radii $a>10$ nm.
 Especially much attention was focused on large nanoparticles of noble metals (gold, silver and copper) because of  location of plasmon resonances in particles of these metals within the visible light spectrum.

An interesting question which we try to discuss in the present paper is the possibility for occurrence of similar plasmon effects with ionic carriers instead of electrons. Many finite ionic systems in a form of enclosed by membranes electrolyte systems can be encountered in biological structures and the question arises regarding possible significance of such ion plasmonic phenomena, the role it would take in such structures and whether the radiative properties of plasmon fluctuation would also be so pronounced in ionic system as they were in metals. It is quite reasonable  that  ionic plasmon effect would be located in other regions of energy and wave lengths in comparison to metallic systems. This, let's call it 'soft' plasmonics could be linked with functionality of biological systems where electricity is rather of ionic than electronic character. For instance the cell signaling, membrane transfer or nerve cell conductivity would serve as examples. 

The theoretical plasmonic model we will adopt for ions, as far as possible, upon analogy to the metallic nanospheres with plasmon excitations theory. The ionic systems are much more complicated in comparison to a  metal crystal structure with free electrons.  Therefore, an  identification  of an appropriate simplifications of the approach to ionic system is of a primary significance. The model must be capable of repetition for ions in electrolyte the plasmonic scenario known from electrons in metals.

In the present paper we will consider the finite spherical ionic system (e.g., liquid electrolyte artificially confined with a membrane) and identify the plasmon excitations of ions in this system. We will determine their energies for various parameters of the ionic system with special attention paid to irradiation properties of ionic plasmons. 

In the subsequent paper \cite{jacak2014b} we will analyze an ionic plasmon-polariton propagation in electrolyte sphere chains with prospective relation to signaling in biological systems. Taking in mind that metallic nano-chains serve as very efficient wave-guides for electro-magnetic signals in the form of collective surface plasmon  excitation of wave-type called plasmon-polaritons,  we will try to model the similar phenomenon in the ionic spheres chains.  

For the initial crude model we will study the spherical or prolate spheroidal ionic conducting system with balanced charges in analogy to the jellium model in metals when local fluctuations of ion density, negative and positive beyond the equilibrium level, can form plasmons in ionic finite system \cite{jacak5}.

\begin{figure}[h]
\centering
\scalebox{0.4}{\includegraphics{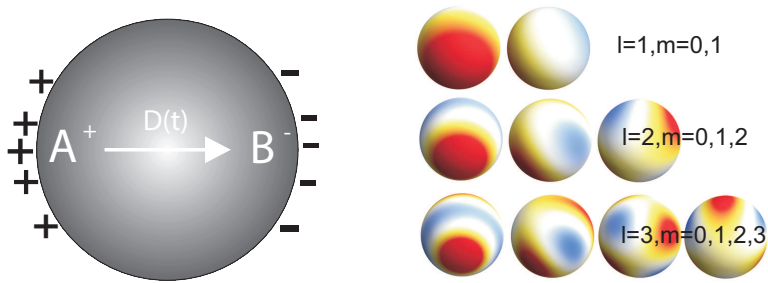}}
\caption{\label{fig1} Dipole $\mathbf{D(t)}$ creation in a single sphere by the simplest surface plasmon oscillations (left); examples of surface plasmon charge distribution with various multiplicity $l,\;m$ ---different colors indicate distinct values of local charge density from negative to positive ones (right) }
\end{figure}

\section{Fluctuations of the charge density in the single conducting ionic spherical system}

The problem of how to establish an adequate model for multi-ionic system to grasp essential properties of ionic plasmons is a main issue. For simple two-component ionic system we deal with both sign ion solutions creating an electrolyte with balanced negative and positive total charge. In equilibrium these charge cancellation is also local.  As we deal with two kinds of carriers they both would form density fluctuations resulting in violation of the local electric equilibrium. The total compensation of both sign charges requires, however, that any density fluctuation of negative charges must be accompanied by distant, in general, but ideally equivalent positive ions fluctuation, and conversely. This means that effectively we deal with density fluctuation of    ions, either positive or negative in charge value (always mutually compensated) with respect to uniform charge distribution assumed as ideally cancelled by the opposite sign uniform background---fictitious jellium in analogy to metal. In this way we can model a two component ionic system by two single component ion systems with jellium of the opposite sign. For simplicity we assume that the charge and mass of the opposite ions are the same, but generalization is straightforward.

To simplify the model according to the above described lines let us consider the spherical shape system with the radius $a$ and with balanced total charge of both sign ions with uniform equilibrium density distributions $n^{+(-)}(\mathbf{r})= n\Theta(a-r)$  ($\Theta(r)$ is the Heaviside step function, $a$ is the sphere radius). For simplicity, we assume the same absolute value of charges of plus- and minus-charged ions. By $m^{+(-)}$ one can denote the mass of positive (negative) ions. They both are of the  order of $10^{4-5} m_e$, where $m_e=3.1 \times 10^{-31}$ kg, is the mass of the electron.  Reducing the two component system to two systems with the jellium, is an approximation but may serve for a recognition of ionic dynamics and of scales for its quantitative characteristics, at least. The advantage of such an approach is a close analogy to description of plasmon in metals including direct definition of the rigid shape of the system by the explicit jellium form. 

Upon the above model assumptions we will consider  the ionic carries with density oscillating around the zero valued equilibrium  density as screened  by  the fictitious positive charged background (the effect of the opposite sign ions presence). Hence, the description of fluctuations of  local density  of electrons in metallic nanosphere can be directly used to model fluctuations of effective ions density, substituting electron mass by the ion  mass and the electron charge by the ion charge. The dynamical equation for the  charged fluid in ion system can be thus repeated  from the case of metal nanosphere with electrons \cite{jacak5}. The equilibrium density of the effective charged liquid, denoted by $n$, will be treated as a parameter and assumed equal to  $n=\eta N_0$  where $\eta$ is the molarity of the electrolyte in the sphere  and   $ N_0$ is the one-molar electrolyte  concentration of ions. The equilibrium density determines the bulk plasmon frequency for the ion system  according  to the formula, $\omega_p^2=\frac{4 \pi n q^2}{m}$, where $n$ and $m$ are the equilibrium uniform concentration and the mass of  ions with the charge $q$. Due to larger $m$ than the electron mass, $m_e$, and usually smaller concentration  of effective ions than the one for electrons in metals,  $\omega_p$ can be considerably reduced, even  by several orders of  magnitude. Note, that for electrons in metals $\hbar \omega_p \simeq 10 $ eV and typically falls into ultra-violet region of radiation with corresponding energy of photons. In the ionic system, the plasmon frequency  $\hbar \omega_p$ can be much lower and placed in the range of  infra-red or even in lower energy part of the electro-magnetic wave spectrum. 

\subsection{Definition of the model}

The  Hamiltonian for the two type ion system  has the form, 
\begin{equation}
\label{hamiltonian}
 \hat{H}_{ion} =
 -\sum\limits_{i =1}^{N^-}\frac{\hbar^2 \nabla_{i}^2}{2m^-} -
\sum\limits_{j =1}^{N^+}\frac{\hbar^2 \nabla_{j}^2}{2m^+}
- \sum\limits_{i,j}^{N^-,N^+} \frac{q^-q^+ }{ \varepsilon |{\bbm r}_i 
  -  {\bbm r}_j|}
 +\frac{1}{2}\sum\limits_{i,i',i\neq i'}^{N^-}\frac{(q^-)^2}{\varepsilon |{\bbm r}_{i}-{\bbm r}_{i'}|} +\frac{1}{2}\sum\limits_{j,j'j\neq j'}^{N^+}\frac{(q^+)^2}{\varepsilon |{\bbm r}_{j}-{\bbm r}_{j'}|},
\end{equation} 
where $q^{-(+)}$,  $M^{-(+)}$,  $N^{-(+)}$ are the charge, the mass and the number  of the $-(+)$ ions, respectively.
To analyze this complicated system we propose the following approximation, assuming, for simplicity,  $q^-=-q^+=q$,  $N^-=N^+=N$,  $m^-=m^+=m$ and let us   add and subtract the same terms as written below,
\begin{equation}
\begin{array}{l}
 \hat{H}_{ion} =
 -\sum\limits_{i =1}^{N}\frac{\hbar^2 \nabla_{i}^2}{2m} -
\sum\limits_{j =1}^{N}\frac{\hbar^2 \nabla_{j}^2}{2m}
- \sum\limits_{i,j} \frac{q^2 }{ \varepsilon |{\bbm r}_i 
  -  {\bbm r}_j|}
 +\frac{1}{2}\sum\limits_{i,i',i\neq i'}^{N}\frac{q^2}{\varepsilon |{\bbm r}_{i}-{\bbm r}_{i'}|} +\frac{1}{2}\sum\limits_{j,j'j\neq j'}^{N}\frac{q^2}{\varepsilon |{\bbm r}_{j}-{\bbm r}_{j'}|}\\
-q^2\sum\limits_{j} \int\frac{n (\mathbf{r})d^3 \mathbf{r}}{\varepsilon|\mathbf{r}_{j}-\mathbf{r}|}
-q^2\sum\limits_{i} \int\frac{n (\mathbf{r})d^3 \mathbf{r}}{\varepsilon|\mathbf{r}_{i}-\mathbf{r}|}
+q^2\sum\limits_{j} \int\frac{n (\mathbf{r})d^3 \mathbf{r}}{\varepsilon|\mathbf{r}_{j}-\mathbf{r}|}
+q^2\sum\limits_{i} \int\frac{n (\mathbf{r})d^3 \mathbf{r}}{\varepsilon|\mathbf{r}_{i}-\mathbf{r}|},\\
\end{array}
\end{equation} 
where we have introduced formally the jellium of the spherical shape for both types of ions, with the density $n$ ideally compensating opposite charges of uniformly distributed ions,  $n({\bbm r})=n\Theta(a-r)$, $a$ is the sphere radius (the positive jellium with the negative jellium mutually cancel themselves).
Assuming that $q^2\sum\limits_{j} \int\frac{n (\mathbf{r})d^3 \mathbf{r}}{\varepsilon|\mathbf{r}_{j}-\mathbf{r}|}
+q^2\sum\limits_{i} \int\frac{n (\mathbf{r})d^3 \mathbf{r}}{\varepsilon|\mathbf{r}_{i}-\mathbf{r}|}- \sum\limits_{i,j} \frac{q^2 }{ \varepsilon |{\bbm r}_i 
  -  {\bbm r}_j|}\simeq 0$, which is fulfilled for not too strong  ion concentration fluctuations beyond the uniform distribution, we can separate the Hamiltonian into the sum, $
\hat{H}_{ions}=\hat{H}^-+\hat{H}^+  $,
 where,
\begin{equation}
\label{hhh}
 \hat{H}^{-(+)} = \sum\limits_{j } \left[  -\frac{\hbar^2 \nabla_{j}^2}{2m} -q^2 \int\frac{n (\mathbf{r})d^3 \mathbf{r}}{\varepsilon |\mathbf{r}_{j}   - \mathbf{r}|}  \right]
 +\frac{1}{2}\sum\limits_{j\neq j'}\frac{q^2}{\varepsilon|\mathbf{r}_{j}-\mathbf{r}_{j'}|}.
\end{equation}
 The latter term in r.h.s. of Eq. (\ref{hhh}) corresponds to interaction between ions of the same sign, whereas the second term in the first sum describes interaction of these ions with the effective jellium (of opposite sign), $\varepsilon $ is the dielectric constant of the electrolyte medium. The confined electrolyte system could be, in particular, a electrolyte medium shaped by appropriately formed  membrane, as frequently occurs in biological systems. Because of separation of the Hamiltonian (\ref{hamiltonian}) one can consider single Hamiltonian (\ref{hhh}). 
The ion  wave function corresponding to Hamiltonian (\ref{hhh})  is denoted by $ \Psi_{ion}(t)$.

The form of Hamiltonian (\ref{hhh}) allows for repetition of its further discussing along the scheme applied to electrons in metals \cite{jacak5}, which we will recall below, for the sake of completeness. A local   density of chosen type ions can be written, in analogy to semiclassical Pines-Bhom random phase approximation (RPA) approach to  electrons in metal \cite{rpa,pines}, in the following form:
\begin{equation}
\rho({\bbm r}, t)=<\Psi_{ion}(t)|\sum\limits_j \delta({\bbm r}-{\bbm r}_j) |\Psi_{ion}(t)>,
\end{equation}
where ${\bbm r}_j$ denotes coordinate of $j-th$  ion and the Dirac delta quasiclassically fixes $j-th$ ion position. The Fourier picture of the above density has the form:
\begin{equation}
\tilde{\rho}({\bbm k}, t)=\int \rho({\bbm r},t) e^{-i{\bbm k}\cdot {\bbm r}} d^3 r = <\Psi_{ion}(t)|\hat{\rho}({\bbm k})|\Psi_{ion}(t)>,
\end{equation}
where the 'operator' $ \hat{\rho} ({\bbm k})=\sum\limits_{j}  e^{-i{\bbm k}\cdot {\bbm r_j}} $.

Using  the above notation one can rewrite $\hat{H}_{ion}$ in the following form, in an analogy to the bulk case for metallic plasmon description \cite{rpa,pines,jacak5}:
\begin{equation}
\label{h1}
\hat{H}_{ion} = \sum\limits_{j =1}^{N} \left[ -\frac{\hbar^2 \nabla_{j}^2}{2m}\right] -
\frac{q^{'2}}{(2 \pi)^3} \int d^3 k \tilde{n}({\bbm k}) \frac{2 \pi}{k^2} \left(\hat{\rho^{+}}({\bbm k}) +  \hat{\rho}({\bbm k})\right)
 + \frac{q^{'2}}{(2 \pi)^3}\int d^3 k \frac{2 \pi}{k^2}\left[ \hat{\rho^{+}}({\bbm k})  \hat{\rho}({\bbm k}) -N \right],
\end{equation}
where:
$ \tilde{n}({\bbm k})=\int d^3 r n ({\bbm r}) e^{-i{\bbm k}\cdot {\bbm r}}$ is the Fourier picture of jellium distribution (in the derivation of Eq. (\ref{h1}) we have taken into account that $  \frac{4 \pi}{k^2}= \int d^3 r \frac{1}{r} e^{-i{\bbm k}\cdot {\bbm r}}$), $q^{'2}=\frac{q^2}{\varepsilon }$ .

Utilizing this form of the effective ion Hamiltonian one can write out the dynamic equation in Heisenberg representation for ion density fluctuations,
\begin{equation}
\frac{d^2 \hat{\rho} ({\bbm k}) }{dt^2}=\frac{1}{(i\hbar)^2} \left[\left[ \hat{\rho} ({\bbm k}),\hat{H}_{ion} \right],\hat{H}_{ion} \right],
\end{equation}
which attains the following form,
\begin{equation}
\label{456}
\begin{array}{l}
\frac{d^2 \hat{\rho} ({\bbm k}) }{dt^2}= -\sum\limits_{j}e^{-i{\bbm k}\cdot {\bbm r}_j}\left\{ -\frac{\hbar^2}{m^2}\left( {\bbm k}\cdot \nabla_j \right)^2
+ \frac{\hbar^2 k^2}{m^2}i  {\bbm k}\cdot \nabla_j +\frac{\hbar^2 k^4}{4 m^2}\right\}\\
-\frac{4\pi q^{'2}}{m (2\pi )^3}\int d^3p  \tilde{n} ({\bbm p})\frac{{\bbm k}\cdot {\bbm p}}{p^2}  \hat{\rho}({\bbm k}- {\bbm p})
-\frac{4\pi q^{'2}}{m (2\pi )^3}\int d^3p  \hat{\rho}({\bbm k}- {\bbm p})\frac{{\bbm k}\cdot {\bbm p}}{p^2}  \hat{\rho}( {\bbm p}).\\
\end{array}
\end{equation}
One can notice that $\hat{\rho}({\bbm k}- {\bbm p}) \hat{\rho}( {\bbm p})=
\delta\hat{\rho}({\bbm k}- {\bbm p}) \delta\hat{\rho}( {\bbm p}) + \tilde{n}({\bbm k}- {\bbm p}) \delta\hat{\rho}( {\bbm p})
+ \delta\hat{\rho}({\bbm k}- {\bbm p}) \tilde{n}( {\bbm p}) +  \tilde{n}({\bbm k}- {\bbm p}) \tilde{n}( {\bbm p})$ and
$  \tilde{n} ({\bbm p})\hat{\rho}({\bbm k}- {\bbm p}) =  \tilde{n} ({\bbm p})\delta\hat{\rho}({\bbm k}- {\bbm p})
+   \tilde{n}({\bbm p})\tilde{n}({\bbm k}- {\bbm p})$, where    $\delta \hat{\rho}({\bbm k}) =  \hat{\rho}({\bbm k)} -  \tilde{n} ({\bbm k})$
describes the 'operator' of
 local ion density fluctuation above the uniform distribution. Therefore, one can rewrite Eq. (\ref{456})
 as follows,
\begin{equation}
\begin{array}{l}
\frac{d^2 \delta \hat{\rho} ({\bbm k}) }{dt^2}=-\sum\limits_{j}e^{-i{\bbm k}\cdot {\bbm r}_j}\left\{ -\frac{\hbar^2}{m^2}\left( {\bbm k}\cdot \nabla_j \right)^2
+ \frac{\hbar^2 k^2}{m^2}i  {\bbm k}\cdot \nabla_j +\frac{\hbar^2 k^4}{4 m^2}\right\}\\
-\frac{4\pi q^{'2}}{m (2\pi )^3}\int d^3p  \tilde{n} ({\bbm k}-{\bbm p})\frac{{\bbm k}\cdot {\bbm p}}{p^2} \delta \hat{\rho}( {\bbm p})
-\frac{4\pi q^{'2}}{m (2\pi )^3}\int d^3p \delta \hat{\rho}({\bbm k}- {\bbm p})\frac{{\bbm k}\cdot {\bbm p}}{p^2}  \delta \hat{\rho}( {\bbm p}).\\
\end{array}
\end{equation}
Taking averaging over the quantum states $|\Psi_{ion}>$, for the ion  density fluctuation
$\delta \tilde{\rho}({\bbm k},t)=   <\Psi_{ion}|\delta\hat{\rho}({\bbm k},t)|\Psi_{ion}>=  \tilde{\rho}({\bbm k},t) -  \tilde{n} ({\bbm k})$, we obtain the following equation,
\begin{equation}
\label{e10}
\begin{array}{l}
\frac{\partial^2 \delta \tilde{\rho} ({\bbm k},t) }{\partial t^2}=-<\Psi_{ion}|\sum\limits_{j}e^{-i{\bbm k}\cdot {\bbm r}_j}\left\{ -\frac{\hbar^2}{m^2}\left( {\bbm k}\cdot \nabla_j \right)^2
+ \frac{\hbar^2 k^2}{m^2}i  {\bbm k}\cdot \nabla_j +\frac{\hbar^2 k^4}{4 m^2}\right\}|\Psi_{ion}>\\
-\frac{4\pi q^{'2}}{m (2\pi )^3}\int d^3p  \tilde{n} ({\bbm k}-{\bbm p})\frac{{\bbm k}\cdot {\bbm p}}{p^2} \delta \tilde{\rho}( {\bbm p},t)
-\frac{4\pi q^{'2}}{m (2\pi )^3}\int d^3p \frac{{\bbm k}\cdot {\bbm p}}{p^2} <\Psi_{ion}| \delta \hat{\rho}({\bbm k}- {\bbm p})  \delta \hat{\rho}( {\bbm p})    |\Psi_{ion}>.\\
\end{array}
\end{equation}

For small $k$,  in analogy to  the semiclassical approximation for electrons \cite{jacak5,pines}, the contributions of the second and third components to the first term on the right-hand-side of Eq. (\ref{e10}) can be neglected as small 
 in comparison to the first component. Small and thus negligible  is also
the  third term  in the right-hand-side of Eq.(\ref{e10}), as involving a product of two $ \delta \tilde{\rho}$ (which we assumed small, $ \delta \tilde{\rho}/n << 1$).
This approach corresponds to the random-phase-approximation (RPA) formulated for bulk metal \cite{rpa,pines}.
Within the RPA,    Eq. (\ref{e10})  attains the following shape,
\begin{equation}
\label{e11}
\frac{\partial^2 \delta \tilde{\rho} ({\bbm k},t) }{\partial t^2}=\frac{2 k^2}{3m}
<\Psi_{ion}|\sum\limits_{j}e^{-i{\bbm k}\cdot {\bbm r}_j}\frac{\hbar^2\nabla_j^2}{2m}|\Psi_{vion}>
-\frac{4\pi q^{'2}}{m (2\pi )^3}\int d^3p  \tilde{n} ({\bbm k}-{\bbm p})\frac{{\bbm k}\cdot {\bbm p}}{p^2} \delta \tilde{\rho}( {\bbm p},t),
\end{equation}
and due to  the spherical symmetry,
$$ <\Psi_{ion}|\sum\limits_{j}e^{-i{\bbm k}\cdot {\bbm r}_j}\frac{\hbar^2}{m^2}\left( {\bbm k}\cdot \nabla_j \right)^2|\Psi_{ion}>
\simeq \frac{2 k^2}{3m}<\Psi_{ion}|\sum\limits_{j}e^{-i{\bbm k}\cdot {\bbm r}_j}\frac{\hbar^2\nabla_j^2}{2m}|\Psi_{ion}>. $$
One can rewrite Eq. (\ref{e11})  in the position representation,
\begin{equation}
\label{e12}
\begin{array}{l}
\frac{\partial^2 \delta \tilde{\rho} ({\bbm r},t) }{\partial t^2}=-\frac{2 }{3m} \nabla^2
<\Psi_{ion}|\sum\limits_{j}\delta({\bbm r}-{\bbm r}_j)\frac{\hbar^2\nabla_j^2}{2m}|\Psi_{ion}>\\
+\frac{\omega_p^2}{4\pi} \nabla \left\{ \Theta(a-r) \nabla \int d^3r_1 \frac{1}{|{\bbm r}-{\bbm r}_1|} \delta \tilde{\rho}( {\bbm r}_1,t)\right\}.\\
\end{array}
\end{equation}
In the case of metals it was next used the Thomas-Fermi formula to assess the   averaged kinetic energy  \cite{rpa}:
\begin{equation}
\begin{array}{l}
<\Psi_{ion}|-\sum\limits_{j}\delta({\bbm r}-{\bbm r}_j)\frac{\hbar^2\nabla_j^2}{2m}|\Psi_{ion}>\simeq
\frac{3}{5} (3\pi^2)^{2/3} \frac{\hbar^2}{2m} (\rho({\bbm r},t))^{5/3}\\
=\frac{3}{5} (3\pi^2)^{2/3} \frac{\hbar^2}{2m}n^{5/3} \Theta(a-r)\left[1+\frac{5}{3}\frac{\delta \tilde{\rho}({\bbm r},t)}{n}+...\right].\\
\end{array}
\end{equation}
The above Thomas-Fermi formula is addressed, however, to fermionic and degenerate quantum systems, as electrons in metals. For ionic systems such estimation of kinetic energy  is inappropriate, because the ion concentration is usually much lower than that one of  electrons in metals and the system is not degenerated even if ions are fermions. The Maxwell-Boltzmann distribution should be applied instead of the Fermi-Dirac or Bose-Einstein ones. Independently of fermionic of bosonic statistics of ions, the Maxwell-Boltzmann distribution allows for estimation of the averaged kinetic energy of ions located inside the sphere with the radius $a$, in the following form,
\begin{equation}
\label{energia}
<\Psi_{ion}|-\sum\limits_{j}\delta({\bbm r}-{\bbm r}_j)\frac{\hbar^2\nabla_j^2}{2m}|\Psi_{ion}>\simeq (n+\delta \rho(\mathbf{r},t))\Theta(a-r) \frac{3kT}{2},    
\end{equation}
where $k$ is the Boltzmann constant and $T$ is the temperature. For ions of the 3D shape  or of the linear shape, the  inclusion of  rotational degrees of freedom would result in the factor $\frac{6kT}{2}$ or $\frac{5kT}{2}$, respectively, instead of $\frac{3kT}{2}$ for point like ion model.

Using the formula (\ref{energia}) and taking  into account that $\nabla \Theta(a-r)=-\frac{\bbm r}{r}\delta(a-r)$,
one can rewrite Eq. (\ref{e12}) in the following manner,
\begin{equation}
\label{e15}
\begin{array}{l}
\frac{\partial^2 \delta \tilde{\rho} ({\bbm r},t) }{\partial t^2}=\left[ \frac{kT}{m}\nabla^2 \delta \tilde{\rho}( {\bbm r},t)-
\omega_p^2 \delta \tilde{\rho}( {\bbm r},t)\right]\Theta(a-r)\\
- \frac{kT}{m} \nabla\left\{\left[n+\delta \tilde{\rho}( {\bbm r},t)\right]\frac{\bbm r}{r}\delta(a-r)
\right\}\\
-\left[\frac{kT}{m}\frac{\bbm r}{r}\nabla \delta \tilde{\rho}( {\bbm r},t)
+      \frac{\omega_p^2}{4\pi}          \frac{\bbm r}{r}\nabla \int d^3r_1 \frac{1}{|{\bbm r}-{\bbm r}_1|} \delta \tilde{\rho}( {\bbm r}_1 ,t)\right]
\delta(a-r).\\
\end{array}
\end{equation}
In the above formula  $\omega_p$ is the bulk ion-plasmon frequency,  $\omega_p^2=\frac{4\pi n q^{'2}}{m}$.
The solution of Eq. (\ref{e15}) can be decomposed into two parts  related   to  the distinct domains---inside the sphere and on the sphere surface,
\begin{equation}
 \delta \tilde{\rho}( {\bbm r,t})=\left\{
           \begin{array}{l}
             \delta \tilde{\rho}_1( {\bbm r,t}), \;for\; r<a,\\
              \delta \tilde{\rho}_2( {\bbm r,t}), \;for\; r\geq a,\; ( r\rightarrow a+),\\
          \end{array}
       \right.
       \end{equation}
corresponding to the volume and surface  excitations, respectively. These two parts of ion local density fluctuations
satisfy the equations (according to Eq. (\ref{e15})),
\begin{equation}
\label{ee20}
\frac{\partial^2 \delta \tilde{\rho}_1 ({\bbm r},t) }{\partial t^2}=\frac{kT}{m} \nabla^2 \delta \tilde{\rho}_1( {\bbm r},t)-
\omega_p^2 \delta \tilde{\rho}_1( {\bbm r},t),
\end{equation}
and (here $\epsilon=0+$)
\begin{equation}
\label{ee21}
\begin{array}{l}
\frac{\partial^2 \delta \tilde{\rho}_2 ({\bbm r},t) }{\partial t^2} =-
\frac{kT}{m} \nabla\left\{\left[ n+\delta \tilde{\rho}_2( {\bbm r},t)\right]\frac{\bbm r}{r}\delta(a+\epsilon -r)\right\}\\
 -  \left[\frac{kT}{m} \frac{\epsilon_F}{m}\frac{\bbm r}{r}\nabla \delta \tilde{\rho}_2( {\bbm r},t)
+      \frac{\omega_p^2}{4\pi}          \frac{\bbm r}{r}\nabla \int d^3r_1 \frac{1}{|{\bbm r}-{\bbm r}_1|} \left(\delta \tilde{\rho}_1( {\bbm r}_1 ,t)
\Theta(a-r_1)+\delta \tilde{\rho}_2( {\bbm r}_1 ,t)\Theta(r_1-a)\right)\right]\delta(a+\epsilon-r).\\
\end{array}
\end{equation}
The Dirac delta  in Eq. (\ref{ee21}) results due to the derivative of the
 Heaviside step function---ideal jellium charge distribution. In Eq. (\ref{ee21}) an infinitesimal shift, $\epsilon =0+$, is introduced
 to fulfill requirements of the Dirac delta definition (its singular point must be an inner point of an open subset of the domain). This shift
 is only of a formal character and does not reflect any asymmetry.

  The electric field due to surface charges  is zero inside the sphere, and therefore cannot influence
  the volume excitations. Oppositely, the volume charge fluctuation-induced-electric-field can excite the surface
  fluctuations.  Therefore, the equation for volume plasmons is  independent of surface plasmons, whereas the volume plasmons contribute the equation for the surface plasmons.

The problem of separation between surface and volume plasmons has been thoroughly analyzed for metal clusters and was identified as significant  for very small clusters. In the size-scale of $1-3$ nm for metallic clusters, the effect of so-called spill-out of electrons beyond the jellium edge was important and caused  the surface fuzzy  resulting in coupling  of volume
         and surface plasmon oscillations. Many  direct numerical simulations (TDLDA, i.e., the time dependent local density approximation) \cite{brack,ekardt} have been verified that the volume--surface excitation miss-mass gradually disappears in larger clusters  \cite{brack,ekardt}, which supports accuracy of semiclassical RPA description, within which 
 volume plasmons can be separated  from the
  surface ones (even though  the latter
 can be excited by the former ones, due to the last term in Eq. (\ref{ee21})).  
The quantum spill-out effect disappears gradually with growing sphere dimension and in the range of several nanometers  for the metallic sphere radius is completely negligible. In the present paper we consider the
       radius range of ionic systems  of micrometer  order, when  quantum effects are negligible.
Such an opportunity  allows us to formulate an analytical
  RPA semiclassical description in the form of an oscillator equation, allowing for phenomenological
   inclusion of the damping effects. The energy dissipation effects turned out to be overwhelming physical property in the case of larger metallic nanospheres \cite{jacak5,jacak11} (with $a>10$ nm for Au or Ag)  and  also  for much larger ionic systems, as we will demonstrate it below. 

\subsection{Solution of RPA equations: volume and surface plasmon frequencies}

\begin{figure}[h]
\centering
\scalebox{0.55}{\includegraphics{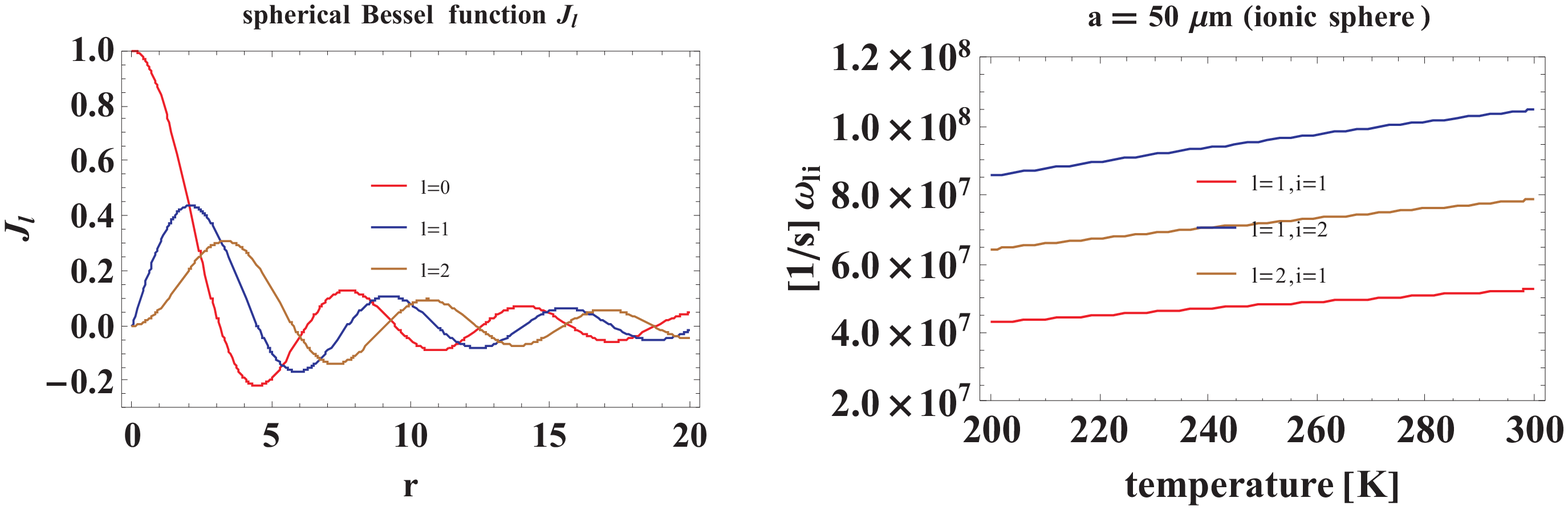}}
\caption{\label{figbes} The spherical Bessel functions $J_l(r)$ for $l=0,1,2$ 
displaying possible charge density fluctuations in the sphere along the sphere radius $r$ (arbitrary units) for volume plasmon modes; the angular distribution for these modes is governed by the real spherical functions $Y_{lm}(\Omega)$ similarly as for the surface plasmon modes (cf. Fig. \ref{fig1} right). The exemplary temperature dependence of self-frequencies of volume plasmon modes $\omega_{li}$, $li=11,\;12,\;21$, for diluted electrolyte $n\simeq 10^{14} $ 1/m$^3$ and ion mass $\sim 10^4 m_e$, $a \sim 50$ $\mu$m---right}
\end{figure}

Eqs (\ref{ee20}) and (\ref{ee21}) are  solved for metallic nanospheres \cite{jacak5} and these solutions can be directly applied to ionic systems.  
To summarize briefly this analysis,  both parts of the plasma fluctuation can represented as   follows,
\begin{equation}
\label{eee111}
           \begin{array}{l}
             \delta \tilde{\rho}_1( {\bbm r,t})=nF({\bbm r}, t), \;for\; r<a,\\
              \delta \tilde{\rho}_2( {\bbm r,t})=\sigma(\Omega,t)\delta(r+\epsilon -a),\;\epsilon=0+,
           \;for\; r\geq a,\; ( r\rightarrow a+),\\
          \end{array}
       \end{equation}
with initial conditions, $  F({\bbm r}, t)|_{t=0}=0, \; \sigma(\Omega,t)|_{t=0}=0$, ($\Omega$ is the spherical angle), $ F({\bbm r}, t)|_{r=a}=0$, $\int\rho({\bbm r},t)d^3r=N$ (neutrality condition). 
For the above initial and boundary conditions and taking advantage of the spherical symmetry,  one can write out  the time-dependent parts of the ion concentration fluctuations in the form \cite{jacak5} (cf. Appendix),
\begin{equation}
\label{e2001}
F({\bbm r}, t) =\sum\limits_{l=1}^{\infty}\sum\limits_{m=-l}^{l}\sum\limits_{i=1}^{\infty}A_{lmn}j_{l}(k_{nl}r)Y_{lm}
(\Omega)sin(\omega_{li}t),
\end{equation}
and
\begin{equation}
\label{e25}
\begin{array}{l}
\sigma(\Omega,t)   = \sum\limits_{l=1}^{\infty}\sum\limits_{m=-l}^{l}  \frac{B_{lm}}{a^2}Y_{lm}(\Omega)sin(\omega_{0l}t)\\
+  \sum\limits_{l=1}^{\infty}\sum\limits_{m=-l}^{l}\sum\limits_{i=1}^{\infty}
A_{lmn}\frac{(l+1)\omega_p^2}{l\omega_p^2-(2l+1)\omega_{li}^2}Y_{lm}(\Omega)n_e\int\limits_0^a dr_1 \frac{r_1^{l+2}}{a^{l+2}}j_{l}(k_{li}r_1)sin(\omega_{li}t),\\
\end{array}
\end{equation}
where $j_l(\xi)=\sqrt{\frac{\pi}{2\xi}}I_{l+1/2}(\xi)$ is the spherical Bessel function, $Y_{lm}(\Omega)$ is the spherical function, $\omega_{li}=
\omega_p\sqrt{1+\frac{kT x_{li}^2}{\omega_p^2a^2m}}$ are the frequencies of the ion volume self-oscillations (volume plasmon frequencies),
$x_{li}$ are the nodes of the Bessel function $j_l(\xi)$ numerated with $i=1,2,3\dots$ (cf. Fig. \ref{figbes}),  $k_{li}=x_{li}/a$,
 $\omega_{l0}=\omega_p\sqrt{\frac{l}{2l+1}}$ are the frequencies of the  ion surface self-oscillations (surface plasmon frequencies). The derivation of the self-frequencies for ionic plasmon oscillations is presented with all the details in Appendix. Amplitudes $A_{lmi}$ and $B_{lm}$ are arbitrary in the homogeneous problem and can be  adjusted to the initial conditions for the first derivatives. 

 The function $F({\bbm r},t)$ describes volume plasmon oscillations, whereas
  $ \sigma(\Omega,t)$ describes the surface plasmon oscillations. Let us emphasize that
  the first term  in the  Eq. (\ref{e25}) corresponds to the surface self-oscillations,
  while the second one term describes the surface  oscillations induced by the volume plasmons.
  The frequencies of the surface self-oscillations are equal to,
  \begin{equation}
  \omega_{0l}=\omega_p\sqrt{\frac{l}{2l+1}},
  \end{equation}
  which, for $l=1$, is the dipole type surface oscillation frequency, described for metallic nanosphere by Mie \cite{mie}, $\omega_{01}=\omega_p/\sqrt{3}$.

\subsection{Ionic surface plasmon frequencies for a nanosphere embedded in a dielectric medium, with $\varepsilon_1 > 1$}

One can now include  the influence of a dielectric surroundings (in general, distinct from the inner one of considered ionic system)  on  plasmons in this system. In order to do it let us assume that  ions on the surface ($r=a+$, i.e., $r\geq a,\;r\rightarrow a$)
interact with Coulomb forces renormalized  by the relative dielectric constant
$\varepsilon_1 > 1$ (distinct from $\varepsilon$ for inner medium). Thus a small modification  of Eq. ({\ref{ee21})  is of order,
\begin{equation}
\label{e26}
\begin{array}{l}
\frac{\partial^2 \delta \tilde{\rho}_2 ({\bbm r}) }{\partial t^2} =-
\frac{2}{3m} \nabla\left\{\left[\frac{3}{5}\epsilon_F n+\epsilon_F \delta \tilde{\rho}_2( {\bbm r},t)\right]\frac{\bbm r}{r}\delta(a+\epsilon -r)\right\}\\
 -  \left[\frac{2}{3} \frac{\epsilon_F}{m}\frac{\bbm r}{r}\nabla \delta \tilde{\rho}_2( {\bbm r},t)
+      \frac{\omega_p^2}{4\pi}          \frac{\bbm r}{r}\nabla \int d^3r_1 \frac{1}{|{\bbm r}-{\bbm r}_1|} \left(\delta \tilde{\rho}_1( {\bbm r}_1 ,t)
\Theta(a-r_1)+\frac{1}{\varepsilon_1}\delta \tilde{\rho}_2( {\bbm r}_1 ,t)\Theta(r_1-a)\right)\right]\delta(a+\epsilon -r),\\
\end{array}
\end{equation}
(note that Eq. (\ref{ee20}) is not affected by the outer medium).
The solution of the above equation is of the same form as that  one for the Eq. (\ref{ee21}) case, but with  the renormalized 
surface plasmon frequencies,
\begin{equation}
 \omega_{0l}=\omega_p\sqrt{\frac{l}{2l+1}\frac{1}{\varepsilon_1}}.
  \end{equation}

\section{Damping of plasmon oscillations in ionic systems}

The semiclassical RPA treatment of plasmon excitations in finite ion systems as presented above, does  not account for plasmon damping. 
The damping of plasmon oscillations  can be, however,  included  in a  phenomenological manner, by addition of an attenuation  term  to  plasmon dynamic equations, i.e., the term,$-\frac{2}{\tau_0}\frac{\partial \delta\rho({\bmb r},t)}{\partial t}$,  added to the right hand sides
  of both Eqs (\ref{ee20}) and (\ref{ee21}), taking into account  their oscillatory form.
The introduced  damping ratio $\frac{1}{\tau_0}$ accounts for ion scattering losses and can be approximated in analogy to metallic systems, by inclusion of energy dissipation caused by irreversible its transformation into heat via various microscopic channels similar to those for Ohmic resistivity  
\cite{atwater},
\begin{equation}
\label{form}
\frac{1}{\tau_0}\simeq \frac{v}{2\lambda_b }+\frac{Cv}{2a},
\end{equation}
where  $a$ is the sphere radius, $v$ is the mean velocity  of ions, $v=\sqrt{\frac{3 k T}{m}}$,
  $\lambda_b$ is the ion mean free path in bulk  electrolyte material the same as the sphere is made of (including  scattering of ions on other ions,
  and on solvent particles and admixtures). The second term in Eq. (\ref{form})  accounts for scattering of ions on the boundary of
    the finite ionic system,  the sphere with the radius $a$, the constant $C$ is of order of unity \cite{atwater}. 

In order to explicitly express a forcing field which moves ions in the system, the inhomogeneous time dependent term should be added to the homogeneous equations (\ref{ee20}) and (\ref{ee21}).
The forcing field  may be the time dependent electric field. If one considers it as the  electric component of the incident e-m wave then the comparison of the  resonant wave-length with the system size is of order. Similarly as for metallic nano-spheres also for finite ionic systems the surface plasmon resonant wave-length highly exceeds the system dimension and the forcing field is practically uniform along whole the system. Such a perturbation could excite only surface dipole plasmons, i.e., the mode with  $l=1$, which can be described   by the function $Q_{1m}(t)$ ($l=1$ and $m$ are angular momentum numbers related to the assumed spherical symmetry of the system).
 The  corresponding dynamical equation for surface plasmons  reduced to only  mode $Q_{1m}(t)$ has  the following form,
  \begin{equation}
  \label{qqq}
  \begin{array}{l}
  \frac{\partial^2Q_{1m}(t)}{\partial t^2}+\frac{2}{\tau_0}\frac{\partial Q_{1m}(t)}{\partial t}+\omega_1^2 Q_{1m}(t)\\
   =\sqrt{\frac{4\pi}{3}}\frac{qn}{m}\left[E_z(t)\delta_{m,0}+\sqrt{2}\left(E_x(t)\delta_{m,1}
   + E_y(t)\delta_{m,-1}\right)\right],\\
   \end{array}
   \end{equation}
   where  $\omega_1=\frac{\omega_p}{\sqrt{3\varepsilon_1}}$ (it is a dipole surface plasmon frequency, i.e., the  Mie frequency \cite{mie}, $\varepsilon_1$ is the dielectric susceptibility of the system surroundings).
 Because  only $Q_{1m}$ contribute to the plasmon response to the homogeneous electric field, thus the effective ion density fluctuation has the form \cite{jacak5},
\begin{equation}
\label{oscyl}
 \delta \rho({\bmb r},t)=\left\{
  \begin{array}{l}
  0,\;\; r<a,\\
\sum\limits_{m=-1}^{1}Q_{1m}(t)Y_{1m}(\Omega)\; r\geq a,\; r\rightarrow a+,\\
  \end{array} \right.
  \end{equation}
 where $Y_{lm}(\Omega)$ is the spherical function with $l=1$.
One can also explicitly calculate the dipol ${\bmb D}(t)$ corresponding to surface  plasmon oscillations given by Eq. (\ref{oscyl}),
\begin{equation}
\label{dipolek}
\left\{
\begin{array}{ll}
D_x(t)&= q'\int d^3r x\delta\rho({\bmb r},t)=  \frac{\sqrt{2\pi}}{\sqrt{3}}q'Q_{1,1}(t)a^3,\\
D_y(t)&= q'\int d^3r y\delta\rho({\bmb r},t)=  \frac{\sqrt{2\pi}}{\sqrt{3}}q'Q_{1,-1}(t)a^3,\\
D_z(t)&= q'\int d^3r z\delta\rho({\bmb r},t)=  \frac{\sqrt{4\pi}}{\sqrt{3}}q'Q_{1,0}(t)a^3.\\
\end{array}\right.
\end{equation}
The dipole ${\bmb D}(t)$ satisfies the equation (it is rewritten  Eq. (\ref{qqq})),
   \begin{equation}
   \label{dipoleq}
   \left[\frac{\partial^2}{\partial t^2}+  \frac{2}{\tau_0}  \frac{\partial}{\partial t} +\omega_1^2\right] {\bmb D}(t)=\frac{a^3 4\pi q^{'2}n}{3m}
   {\bmb E}(t)=\varepsilon a^3 \omega_1^2 {\bmb E}(t).
   \end{equation}

One can notice that  the dipole (\ref{dipolek}) scales as the system volume,  $\sim a^3$, which may be interpreted  that all ions  actually contribute  to the  surface plasmon oscillations. This is connected with the fact that the surface modes correspond to  uniform translation-type oscillations of ions in the system, when inside the sphere the charge of ions is exactly compensated by oppositely  signed ions,  whereas the  not balanced charge density occurs only on the surface despite all ions oscillate. For the volume plasmons the non-compensated charge density fluctuations are present also inside the sphere as volume plasmon modes have the compressional character with not balanced charge fluctuations along the system radius.    

  The scattering effects  accounted for by the approximate formula (\ref{form}) cause  damping of plasmons especially strong
for small size of the system due to the nanosphere-edge scattering contribution
 proportional to $\frac{1}{a}$. This term is, however, 
  of lowering significance with the radius growth.  We will show that radiation losses resulted due to accelerated movement of ions     scales  as $a^3$, and for rising $a$ these irradiative energy losses quickly dominate plasmon attenuation. Due to opposite size dependence of scattering and irradiation contributions to the plasmon damping one can thus observe the   cross-over  in dumping with respect  to its size dependence, as it is depicted in Fig. \ref{lorentz45}. One can also determine the radius $a^*$ for which the total attenuation rate for surface plasmons is minimal, $a^*=\left(\frac{3^{3/2} C c^3 v}{2 \omega_1 \omega_p^3}\right)^{1/4} $. The system sizes $a^*$ for two  distinct ionic systems are listed in Tab. \ref{tab1}.

\begin{figure}
\centering
\scalebox{0.35}{\includegraphics{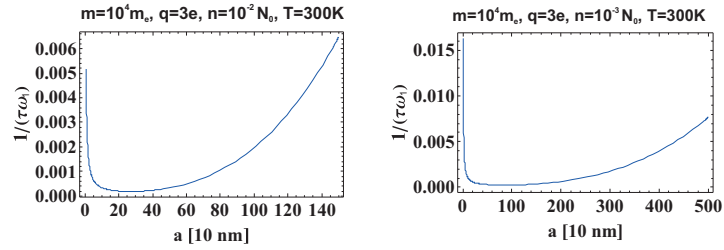}}
\caption{\label{lorentz45} The cross-over in ionic system-size dependence of damping rate for surface plasmons, for $T=300$K, $m=10^4m_e$, $q=3e$, $n=10^{-2}N_0$ ($N_0$ is the concentration of one molar electrolyte) (left) and for $n=10^{-3}N_0$ (right); in the size-region close to the cross-over the perturbative treatment for Lorentz friction perfectly coincides with the exact approach}
\end{figure}

The irradiation of energy of the oscillating  dipole is  expressed by the so-called Lorentz friction \cite{lan}, i.e., the effective electric field slowing down the motion of  charges,
\begin{equation}
\label{lorentzforce}
{\bmb E}_L=\frac{2\sqrt{\varepsilon}}{3 c^3}\frac{\partial^3{\bmb D(t)}}{\partial t^3}.
\end{equation}
Hence, we
can rewrite Eq. (\ref{dipoleq}) including the Lorentz friction term, 
\begin{equation}
\label{dipoleql}
   \left[\frac{\partial^2}{\partial t^2}+  \frac{2}{\tau_0}  \frac{\partial}{\partial t} +\omega_1^2\right] {\bmb D}(t)=\varepsilon a^3 \omega_1^2 {\bmb E}(t)+ \varepsilon a^3 \omega_1^2 {\bmb E}_L,
   \end{equation}
or for ${\bmb E}=0$, 
\begin{equation}
\label{dipoleqlll}
   \left[\frac{\partial^2}{\partial t^2} +\omega_1^2\right] {\bmb D}(t)=    \frac{\partial}{\partial t}\left[ -\frac{2}{\tau_0} {\bmb D}(t) + \frac{2}{3\omega_1}\left(\frac{\omega_p a}{c\sqrt{3}}\right)^3 \frac{\partial^2}{\partial t^2} {\bmb D}(t)  \right] .
   \end{equation} 
One can apply  the perturbation method for solution of Eq. (\ref{dipoleqlll}) 
when  the right hand side of this equation is treated  as a small perturbation. In the zeroth step of the perturbation we have $\left[\frac{\partial^2}{\partial t^2} +\omega_1^2\right] {\bmb D}(t)=    0$, from which $ \frac{\partial^2}{\partial t^2} {\bmb D}(t) = -\omega_1^2 {\bmb D}(t)$. Hence, for the first step of the perturbation, we put the latter formula to the right hand side of Eq. (\ref{dipoleqlll}), i.e.,  
\begin{equation}
\label{dipoleqllllll}
   \left[\frac{\partial^2}{\partial t^2}+ \frac{2}{\tau} \frac{\partial}{\partial t} +\omega_1^2\right] {\bmb D}(t)=    0 ,
   \end{equation}
where 
\begin{equation}
\label{tau111}
\frac{1}{\tau}=\frac{1}{\tau_0} +\frac{\omega_1}{3}\left( \frac{\omega_p a}{c\sqrt{3}}\right)^3.
\end{equation}
Within the first step of perturbation,  the Lorentz friction can be  included into the total attenuation rate $\frac{1}{\tau}$. Nevertheless, this approximation is justified  only for sufficiently small perturbations, i.e., when the second term in Eq. (\ref{tau111}), proportional to  $a^3$, is  small enough  to fulfill the perturbation restrictions. The related limiting value, $\tilde{a}$, of the ionic system size depends  on the ion concentration, charge, mass, dielectric susceptibility, as is exemplified below in the following subsection.

The solution of Eq. (\ref{dipoleqllllll}) is of the form ${\bmb D}(t)=
{\bmb A} e^{-t/\tau}cos(\omega_1' t + \phi)$, where $\omega_1' =\omega_1'\sqrt{1-\frac{1}{(\omega_1\tau)^2}}$, which gives the  red shift of the plasmon resonance due to strong, $\sim a^3$, growth of attenuation caused  by the irradiation. The Lorentz friction  term in Eq. (\ref{tau111}) dominates plasmon damping for $a\leq \tilde{a}$  due to this   $a^3$ dependence---cf. Fig. \ref{lorentz45}. The plasmon damping grows rapidly with $a$ and this results  in pronounced redshift of resonance frequency.

\subsection{Exact inclusion of the Lorentz damping to the attenuation of ionic dipole surface plasmons}

Now we will consider the dynamic equation for surface plasmons in the ionic spherical system (\ref{dipoleqlll}) with the Lorentz friction term, but without application of 
the perturbation method for solution  resulting in substitution of the Lorentz friction term 
$\frac{2}{3\omega_1}\left(\frac{\omega_p a}{v\sqrt{3}}\right)^3 \frac{\partial^3 {\bbm D}(t)}{\partial t^3}$ with the approximate formula, $-\frac{2\omega_1}{3}
\left(\frac{\omega_p a}{v\sqrt{3}}\right)^3 \frac{\partial {\bmb D}(t)}{\partial t}$, what
was the result of taking in the right hand side of Eq. (\ref{dipoleqlll}) the zeroth order its solution, for which $ \frac{\partial^2 {\bmb D}(t)}{\partial t^2}=-\omega_1^2 \mathbf{D}(t)$. 
 To compare various contributions to Eq. (\ref{dipoleqlll}) we change to dimensionless  variable $t\rightarrow t'= \omega_1t$. Then Eq. (\ref{dipoleqlll}) attains the form,
\begin{equation}
\label{dipoleqnnn}
\frac{\partial^2 \mathbf{D}(t')}{\partial {t'}^2}+\frac{2}{\tau_0 \omega_1} \frac{\partial \mathbf{D}(t')}{\partial t'}+\mathbf{D}(t')=\frac{2}{3}\left(\frac{\omega_p a}{v\sqrt{3}}\right)^3\frac{\partial^3 \mathbf{D}(t')}{\partial {t'}^3}.
\end{equation}

In the case of solution of Eq. (\ref{dipoleqnnn}) by perturbation we get renormalized attenuation rate for effective damping term, $\frac{1}{\omega_1 \tau_0}+\frac{1}{3}\left(\frac{\omega_p a}{v \sqrt{3}}\right)^3$. This term quickly achieves the value 1, for which the oscillator falls into the over-damped regime. For system parameters as assumed for Fig. \ref{lorentz45}, the  achievement by the attenuation rate of the value equal 1 takes place at 25,5 $\mu$m  and 8 $\mu$m for $n=10^{-3}N_0$ and $n=10^{-2}N_0$, respectively. At these values of $a$, the frequency $\omega_1' =\omega_1'\sqrt{1-\frac{1}{(\omega_1\tau)^2}}$ goes to zero, which indicates 
an apparent artifact of the perturbation  method. To verify how behaves the exact damped frequency in the considered system one has to solve the dynamical equation without any approximations. As this equation is of third order linear differential equation, one can find its solution in the form, 
 $\sim e^{i \Omega t'}$, with the analytical expressions for three possible values of the exponent,
\begin{equation}
\label{kilka}
\begin{array}{ll}
\Omega_1 =&-\frac{i}{3g}-\frac{i 2^{1/3}(1+6 gu)}{3g\left(2+27 g^2+18 gu + \sqrt{4(-1-6 gu)^3+(2+27 g^2+18 gu)^2}\right)^{1/3}}\\
&-\frac{i \left(2+27 g^2+18 gu +\sqrt{4(-1-6gu)^3+(2+27 g^2+18gu)^2}  \right)^{1/3}}{ 3 \times 2^{1/3}g} \in Im (=i \alpha ),\\
\Omega_2=&-
\frac{i}{3g}+\frac{i(1+ i\sqrt{3})(1+6 gu)}{ 3\times 2^{2/3} g\left(2+27 g^2+18 gu + \sqrt{4(-1-6 gu)^3+(2+27 g^2+18 gu)^2} \right)^{1/3}}\\
&+\frac{i(1-i\sqrt{3}) \left(2+27 g^2+18 gu +\sqrt{4(-1-6gu)^3+(2+27 g^2+18gu)^2}\right)^{1/3}}{ 6 \times 2^{1/3}g}=   \omega  + i \frac{1}{\tau },\\
\Omega_3=&-
\frac{i}{3g}+\frac{i(1- i\sqrt{3})(1+6 gu)}{ 3\times 2^{2/3} g\left(2+27 g^2+18 gu + \sqrt{4(-1-6 gu)^3+(2+27 g^2+18 gu)^2}  \right)^{1/3}}\\
&+\frac{i(1+i\sqrt{3}) \left(2+27 g^2+18 gu +\sqrt{4(-1-6gu)^3+(2+27 g^2+18gu)^2}\right)^{1/3}}{ 6 \times  2^{1/3}g}=-\omega+i \frac{1}{\tau  },\\
\end{array}
\end{equation}
where 
$u = \frac{1}{\tau_0 \omega_1}$ and 
$g = 2/3  \left(\frac{a\omega_p}{c \sqrt{3\varepsilon_1}}\right)^3$.

\begin{figure}[h]
\centering
\scalebox{0.4}{\includegraphics{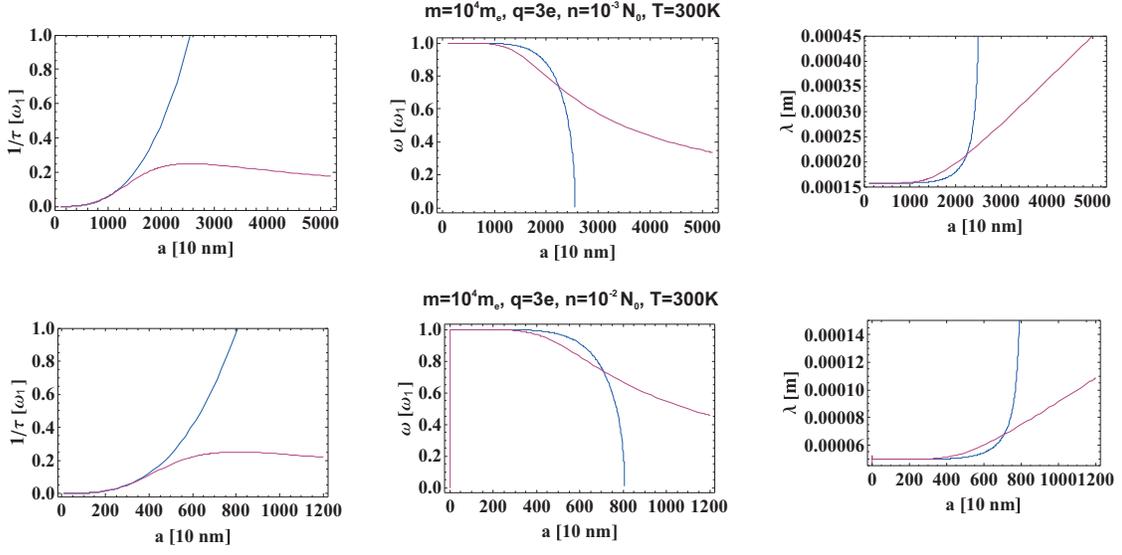}}
\caption{\label{lorentz52}
Comparison of the damping rate and the resonance frequency (transformed also into the resonance wave-length---in right panels), i.e., the damping rate and  frequency (wave length) of oscillating solution of Eq.(\ref{dipoleqnnn}), exact (red line) and approximate upon perturbation approach (blue line), both with respect to the ionic finite system  radius $a$}
\end{figure}

In  Fig. \ref{lorentz52} we have plotted the damping rate ($Im \Omega$) and  the  self-frequency ($Re \Omega$) (also translated for resonance wave-length---right pannels) with respect to the system radius $a$. For comparison,  the approximate perturbative solutions are plotted also---the blue line, whereas   the exact solution of Eq. (\ref{dipoleqnnn}) is plotted in red line. The blue line finishes at $a_{limit}$, when the  attenuation rate within the perturbation approach reaches the critical value 1 (then $\lambda \rightarrow \infty$). For the accurate solution of Eq. (\ref{dipoleqnnn}) this singular behavior disappears and the oscillating solution, $e^{i\Omega t}$,  exists for larger $a$ as well. 

We notice that the red-shift of the plasmon resonance is strongly overestimated in the framework of the perturbative approach to the Lorentz friction unless $a<\tilde{a}$, where $\tilde{a}$ is sensitive to ionic system parameters and especially to ion concentration (as is demonstrated in Fig. \ref{lorentz52}).

Let us emphasize   that the equation (\ref{dipoleqnnn}) has in general two types of particular solutions, $e^{i\Omega t'}$, with complex self-frequencies $\Omega$. The solutions given by $\Omega_2$ and $\Omega_3$ are of oscillating type with damping ($i\Omega_2$ and $i\Omega_3$  are mutually conjugated, thus $\Omega_2$ and $\Omega_3 $ have  the real parts of opposite sign, whereas  the same imaginary parts, the latter is positive displaying  the damping rate) and the second one---given by  $\Omega_1$, which turns out to be an unstable exponentially rising solution (negative imaginary solution). This unstable solution is the well known artifact in the  Maxwell electrodynamics (cf. e.g.,  \$  75 in \cite{lan}) and corresponds  to the infinite self-acceleration of the free charge due to Lorentz friction force (i.e., to the singular solution of the equation $m\dot{\mathbf{v}}=const. \times \ddot{\mathbf{v}}$, which is associated with a formal renormalization  of the field-mass of the charge---infinite for point-like charge and canceled in an artificial manner by arbitrary assumed negative infinite non-field mass, resulting  in ordinary mass of e.g., an electron, which is, however,  not defined mathematically in a proper way). This unphysical singular particular solution should be thus discarded. The other oscillatory type solution resembles the solution of the ordinary damped harmonic oscillator, though with distinct attenuation rate and frequency. They are expressed by analytical formulae for $\Omega_2$ and $\Omega_3$ by Eqs (\ref{kilka}) and then are calculated for various $a$ and compared with the corresponding quantities found within the perturbation approach. This comparison is presented in Fig. \ref{lorentz52}. 
From this comparison it is clearly visible that application of the perturbation approach leads to high overestimation of the damping rate for $a>\tilde{a}$. Therefore, we can conclude that the usage of  the approximate formula for the 
Lorentz friction damping in the form (\ref{tau111}) is justified up to $a \simeq \tilde{a}$, while for $a>\tilde{a}$, these approximate values strongly differ from the exact ones. The value $\tilde{a}<a_{limit}$ sharply depends on ionic system parameters and approximately $\tilde{a}\simeq\frac{a_{limit}}{2}$.

\begin{table}[t]
\centering
\begin{tabular}{p{5cm}|p{4cm}|p{3cm}|p{3cm}}
\hline
material& ionic system & sample 1 & sample 2 \\
\hline
ion concentration&$n$ ($N_0$ is one-molar concentr.)&$10^{-2}\; N_0$&$10^{-3}\;N_0$\\
effective ion mass &$m$ ($m_e$ electron mass)&$10^4\; m_e$&$10^4\; m_e$\\
charge of effective ion &$q/\sqrt{\varepsilon}$& $3\;e$&$3\;e$\\
temperature& $T $& 300 K& 300 K\\
mean velocity of ions&$ v=\sqrt{\frac{3 k T }{ m}}$& 1168 m/s& 1168 m/s\\
bulk plasmon frequency&  $\omega_p$&$9.3 \times 10^{13}$ 1/s    &$2.93 \times 10^{12}$ 1/s \\
dielectric constant of souroundings&$\varepsilon_1$&2&2\\
Mie frequency& $\omega_1=\omega_p/\sqrt{3\varepsilon_1}$&$3.8 \times 10^{13}$ 1/s    &$1.2\times 10^{12}$ 1/s\\
constant in Eq. (\ref{form})& $C$& 2&2\\
bulk mean free path (room temp.) &$\lambda_b$&  $0.5\; \mu$m&$0.1\; \mu$m \\
radius for minimal damping&$a^*=\left(\frac{3^{3/2}Cc^3v}{2\omega_1\omega_p^3}\right)^{1/4}$&$2.7 \times 10^{-7}$ m&$8.6 \times 10^{-7}$ m\\
\hline
\end{tabular}
\caption{The ion-system parameters assumed for calculation of damping rates and self-frequency  for dipole surface  plasmons }
\label{tab1}
\end{table}

\section{Conclusions}

Concluding,  we can state that in ionic finite systems we may observe plasmons similar as in the metallic nanoparticles. The structure of surface and volume plasmons for ions is repeated from the similar properties of electronic plasmons in metallic spherical systems, however, with the significant shift of resonant energy towards lower one correspondingly to by few orders larger mass of ions in comparison to electron mass and different concentration of ions in electrolyte. 
Thus corresponding to the resonant energy electro-magnetic wave length is shifted to deep infra-red or even longer wave lengths  depending on ion concentration. The   typical for metal clusters  cross-over in size dependence of plasmon damping between the  scattering, leading to  Ohmic type energy dissipation,  versus the irradiation losses also is observable in ionic spherical system with similar to metal size-dependence, though shifted toward the micrometer scale for ions instead on the nanometer scale for metals. Of particular interest is the high irradiation regime for dipole plasmons in ionic system with prospective application for signaling and energy transfer in ionic systems. The initial strong enhancement of efficiency of the Lorentz friction with the radius growth of electrolyte sphere is observed on the micrometer scale with typical $a^3$ radius  dependence above some threshold radius which value depends on electrolyte parameters. At certain value of the radius (variating in a wide range also depending on ion system parameters) this enhancement saturates and then the irradiation losses slowly diminish, which allows for definition of the most convenient sizes of electrolyte finite system  for optimizing radiation mediated transport  efficiency preferring the highest radiation losses.  

\vspace{3mm}

{\bf Acknowledgments} Authors 
acknowledge the support of the present work upon the NCN
project\\ no. 2011/03/D/ST3/02643 and  the NCN project no. 2011/02/A/ST3/00116.

\appendix
\section{Derivation of plasmon frequencies}
\subsection{Volume plasmons}
In order to determine self-frequencies of the volume ionic plasmon in the sphere we must solve Eq. (\ref{ee20}) with the form of the relevant solution given by Eq. (\ref{eee111}). For initial conditions listed below Eq. (\ref{eee111}) we assume $F(\mathbf{r},t)=F_{\omega}(\mathbf{r})sin(\omega t)$ and by substitution of this function into Eq. (\ref{ee20}) we get,
\begin{equation}
    \Delta F_{\omega}(\mathbf{r})+k^2F_{\omega}(\mathbf{r})=0,
\end{equation}
where $k^2=\frac{(\omega^2-\omega_p^2)m}{kT}$. This is a well known Helmholtz differential equation  of which solutions (finite in the origin) can be  expressed by the spherical Bessel functions (for the radial dependence of $F_{\omega}(\mathbf{r})$), 
\begin{equation}
    F_{\omega}(\mathbf{r})=Aj_l(kr)Y_{lm}(\Omega),
\end{equation} 
where $j_l(x)=\sqrt{\frac{\pi}{2x}}I_{l+1/2}(x)$ is the the $l$th  spherical Bessel function linked to the Bessel function of first kind. The boundary condition, $F(a)=0$, gives quantization of $k$, $k_{li}=\frac{x_{li}}{a}$, where $x_{li}$ is the $i$th zero of the $l$th Bessel function (cf.   Fig. \ref{figbes} left). Following  this quantization one arrives at the corresponding self-frequency quantization,
\begin{equation}
\label{frv}
    \omega_{li}^2=\omega_p^2\left(1+\frac{kTx_{li}}{\omega_p^2 m a^2}\right).
\end{equation}  
Thus the volume ionic plasmons in the sphere are described by functions,
\begin{equation}
    \delta\rho_1(\mathbf{r},t)=n\sum\limits_{l=1}^{\infty}\sum\limits_{m=-l}^{m=l}\sum_{i=1}^{\infty}A_{lmi}
j_l(k_{li}r)Y_{lm}(\Omega)sin(\omega_{li}t),
\end{equation}
where $A_{lmi}$ are arbitrary constants. The component with $l=0$ vanishes because of neutrality condition, $\int\limits_{0}^{a}r^2 dr d\Omega F(\mathbf{r},t)=0$ (as $\int d\Omega Y_{lm}(\Omega)=\sqrt{4\pi} \delta_{l0}\delta_{m0}$, $d\Omega =sin\Theta d\Theta d\phi$).
Note that in ionic systems self-frequencies of volume plasmons in the sphere are temperature dependent---cf Eq. (\ref{frv}) and Fig. \ref{figbes} right.  

\subsection{Surface plasmons}

In order to determine self-frequencies for surface plasmons, one has to consider Eq. (\ref{ee21}) and solution for it given by Eq. (\ref{eee111}). 
The first term in the right hand side of Eq. (\ref{ee21}) can be rewritten to the form,
\begin{equation}
  \begin{array}{l}  \frac{kT}{m}\nabla(n+\delta\rho_2)\nabla\Theta(a-r)+\frac{kT}{m}(n+\delta\rho)\Delta\Theta\\
=-\frac{kT}{m}\delta(a-r)\frac{\partial}{\partial r}(n+\delta\rho)=\frac{kT}{m}\frac{1}{r^2}\frac{\partial}{\partial r}(r^2\delta(a-r)\\
=-\frac{kT}{m}\frac{1}{r^2}\frac{\partial}{\partial r }\left[(n+\delta \rho_2)r^2\delta(a-r)\right],
\end{array}  
\end{equation}
where we used formulae $\nabla \Theta(a-r)=-\frac{\mathbf{r}}{r}\delta(a-r)$, $\frac{\mathbf{r}}{r} \nabla  =\frac{\partial}{\partial r}$.
The next term in the right hand side of Eq. (\ref{ee21}) can be transformed into,
\begin{equation}
\begin{array}{l}
    -\frac{kT}{m}\delta (a-r)\frac{\mathbf{r}}{r}\nabla \delta\rho_2-\frac{\omega^2_p}{4\pi}\delta(a-r)\frac{\mathbf{r}}{r}\nabla\int \frac{d^3r_1\delta\rho(\mathbf{r_1})}{|\mathbf{r}-\mathbf{r_1}|}\\
=-\frac{kT}{m}\delta(a-r)\frac{\partial }{\partial r}\delta\rho-2-
\frac{\omega^2_p}{4\pi}\delta(a-r) \frac{\partial}{\partial r}\int\frac{d^3r_1
\delta\rho(\mathbf{r}_1)}{|\mathbf{r}-\mathbf{r}_1|}.
\end{array}
\end{equation}
Eq. (\ref{ee21}) attains thus the form,
\begin{equation}
\label{ee211}
\begin{array}{l}
  \frac{\partial^2\rho_2}{\partial t^2}= -\frac{kT}{m}\frac{1}{r^2}\frac{\partial}{\partial r }\left[(n+\delta \rho_2)r^2\delta(a-r)\right]\\ -\frac{kT}{m}\delta(a-r)\frac{\partial }{\partial r}\delta\rho-2-
\frac{\omega^2_p}{4\pi}\delta(a-r) \frac{\partial}{\partial r}\int\frac{d^3r_1
\delta\rho(\mathbf{r}_1)}{|\mathbf{r}-\mathbf{r}_1|}.\\
\end{array}
\end{equation}
We suppose the solution of the above equation in the form, $\delta\rho_2=\sigma(\Omega, t)\delta(a+0^+-r)$ and multiply both sides of this equation by $r^2$ and integrate with respect to $r$ in arbitrary limits, i.e., $\int\limits_l^Lr^2dr\dots$, such that  $a\in(l,L)$ (this integration removes Dirac deltas), which leads to the equation,
\begin{equation}
\label{ee2111}
\begin{array}{l}
    a^2\frac{\partial^2 \sigma(\Omega,t)}{\partial t^2}=-\frac{kT}{m}\int\limits_l^Ldr\frac{\partial}{\partial r}\left[ (n+\delta\rho_2)r^2\delta(a-r)\right]\\
 -\frac{kT}{m}\sigma(\Omega,t)\int\limits_l^Lr^2 dr \delta(a-r)\frac{\partial}{\partial r}\delta(a-r)\\
-\frac{\omega_p^2}{4\pi} \int\limits_l^Lr^2dr \delta(a-r)\frac{\partial }{\partial r}\int\limits_a^{\infty}r_1^2dr_1\int d\Omega\frac{\delta\rho_1(\mathbf{r}_2)}{|\mathbf{r}
-\mathbf{r}_1|}\\
-\frac{\omega_p^2}{4\pi} \int\limits_l^Lr^2dr \delta(a-r)\frac{\partial }{\partial r}\int\limits_0^ar_1^2dr_1\int d\Omega\frac{\delta\rho_1(\mathbf{r}_1)}{|\mathbf{r}
-\mathbf{r}_1|}.\\
\end{array}
\end{equation}
Two first terms in the  right hand side of the above equation vanish, because,
\begin{equation}
  -\frac{kT}{m}\int\limits_l^Ldr\frac{\partial}{\partial r}\left[ (n+\delta\rho_2)r^2\delta(a-r)\right]=-\frac{kT}{m}\left[(n+\delta\rho_2)r^2
\delta(r-a)\right]  |_l^L=0  
\end{equation}
and
\begin{equation}
\begin{array}{l}
 -\frac{kT}{m}\sigma(\Omega,t)\int\limits_l^Lr^2 dr \delta(a-r)\frac{\partial}{\partial r}\delta(a-r)=-\frac{kT}{m}a^2\int\limits_l^Ldr \frac{1}{2}\frac{\partial}{\partial r}\delta^2(a-r)\\
=-\frac{kT}{m}\frac{a^2}{2}\delta^2(a-r)|_l^L=-\frac{kT}{m}\frac{a^2}{2}
lim_{\mu\rightarrow 0}\frac{1}{\pi}\frac{\mu}{\mu^2+(a-r)^2} \delta(a-r)|_l^L=0.\\
\end{array}  
\end{equation}
Two last terms of r.h.s.  of  Eq. (\ref{ee2111}) can be transformed using the formula \cite{gradst}, $\frac{1}{\sqrt{1+z^2-2zcos\gamma}}=\sum\limits_{l=0}^{\infty}P_l(cos\gamma)z^l,\;for\; \;z<1$, where $P_l(cos\gamma)=\frac{4\pi}{2l+1}\sum\limits_{m=-l}^lY_{lm}(\Omega)Y^*_{lm}(\Omega)$ are Legendre polynomials. This formula leads to the following one,
\begin{equation}
\label{for}
    \frac{\partial}{\partial a}\frac{1}{|\mathbf{a}-\mathbf{r}_1|}=\left\{\begin{array}{l}
\sum\limits_{l=0}^{\infty}\frac{la^{l-1}}{r_1^{l+1}}P_l(cos\gamma),\;for\;a<r_1,\\
-\sum\limits_{l=0}^{\infty}\frac{(l+1)r_1^l}{a^{l+2}} P_l(cos\gamma),\;for \;a>r_1,\\
\end{array}\right.
\end{equation}
where $\mathbf{a}=a\frac{\mathbf{r}}{r}$, $cos\gamma=\frac{\mathbf{a}\cdot\mathbf{r}_1}{ar_1}$.
Employing Eq. (\ref{for}), the last two terms in Eq. (\ref{ee2111}) can be transformed as follows,
\begin{equation}
    \begin{array}{l}
-\frac{\omega_p^2}{4\pi}\int\limits_l^Lr^2dr\delta(a-r)\frac{\partial}{\partial r}\int\limits_a^{\infty}r_1^2dr_1\int d\Omega_1\frac{\delta\rho_2(\mathbf{r}_1)}{|\mathbf{r}
-\mathbf{r}_1|}\\
=-\frac{\omega_p^2}{4\pi}a^2\int d\Omega_1 \int\limits_a^{\infty}r_1^2dr_1\delta\rho_2(\mathbf{r}_1)\frac{\partial}{\partial a}\frac{1}{\sqrt{a^2+r^2_1-2ar_1cos\gamma}}\\
=-\frac{\omega_p^2}{4\pi}a^2\int d\Omega_1  \int\limits_a^{\infty} r_1^2dr_1 \sigma(\Omega_1) \delta(a+0+-r_1)    \sum\limits_{l=0}^{\infty}\frac{la^{l-1}}{r_1^{l+1}}P_l(cos\gamma)\\
=-\frac{\omega_p^2}{4\pi} a^2 \int d\Omega_1 \sigma (\Omega_1) \frac{1}{a^2}
\sum\limits_{l=0}^{\infty} \frac{4\pi l}{2l+1}\sum\limits_{m=-l}^l Y_{lm}(\Omega)
Y^*_{lm}(\Omega_1)\\
=-\omega_p^2 a^2\sum\limits_{l=0}^{\infty}\sum\limits_{m=-l}^l\frac{l}{2l+1}Y_{lm}(\Omega)
\int d\Omega_1\sigma(\Omega_1)Y^*_{lm}(\Omega_1),\\
\end{array}
\end{equation}
and 
\begin{equation}
    \begin{array}{l}
-\frac{\omega_p^2}{4\pi}\int\limits_l^Lr^2dr\delta(a-r)\frac{\partial}{\partial r}\int\limits_0^{a}r_1^2dr_1\int d\Omega_1\frac{\delta\rho_1(\mathbf{r}_1)}{|\mathbf{r}
-\mathbf{r}_1|}\\
=-\frac{\omega_p^2}{4\pi}a^2\int d\Omega_1 \int\limits_0^{a}r_1^2dr_1 nF(\mathbf{r}_1,t)(\mathbf{r}_1)\frac{\partial}{\partial a}\frac{1}{\sqrt{a^2+r^2_1-2ar_1cos\gamma}}\\
=\frac{\omega_p^2}{4\pi} a^2 \int d\Omega_1 nF(\mathbf{r_1},t) 
\sum\limits_{l=0}^{\infty} \frac{(l+1)r_1^l}{a^{l+2}}P_l(cos\gamma)\\
=\omega_p^2 n 
\sum\limits_{l=0}^{\infty} \frac{l+1}{2l+1}Y_{lm}(\Omega)\int\limits_0^ar_1^2dr_1\frac{r_1^l}{a^l}\sum\limits_{l_1=1}^{\infty}
\sum\limits_{m_1=-l_1}^{l_1}\sum\limits_iA_{lmi}j_{l_1}(k_{l_1i}r_1)sin(\omega_{l_1i}t)\int d\Omega_1Y^*_{lm}(\Omega_1)Y_{l_1m_1}(\Omega_1)\\
=\omega_p^2 n\sum\limits_{l=0}^{\infty}\sum\limits_{m=-l}^l \sum\limits_{i}\frac{l+1}{2l+1}Y_{lm}(\Omega)
A_{lmi}\int\limits_0^a\frac{r_1^{l+2} dr_1}{a^l}j_l(k_{li}r_1)sin(\omega_{li}t).\\
\end{array}
\end{equation}
Equation (\ref{ee2111}) attains thus the form,
\begin{equation}
\begin{array}{l}
   \frac{\partial^2 \sigma(\Omega,t)}{\partial t^2}
=-\omega_p^2 a^2\sum\limits_{l=0}^{\infty}\sum\limits_{m=-l}^l\frac{l}{2l+1}Y_{lm}(\Omega)
\int d\Omega_1\sigma(\Omega_1)Y^*_{lm}(\Omega_1)\\
+\omega_p^2 n\sum\limits_{l=0}^{\infty}\sum\limits_{m=-l}^l \sum\limits_{i}\frac{l+1}{2l+1}Y_{lm}(\Omega)
A_{lmi}\int\limits_0^a\frac{r_1^{l+2} dr_1}{a^l}j_l(k_{li}r_1)sin(\omega_{li}t),\\
\end{array}
\end{equation}
 Assuming now, $\sigma(\Omega,t)=\sum\limits_{l=0}^{\infty}\sum\limits_{m=-l}^lq_{lm}(t)Y_{lm}(\Omega)$ and putting it to the above equation, we obtain,
\begin{equation}
   \begin{array}{l} \sum\limits_{l=0}^{\infty}\sum\limits_{m=-l}^lY_{lm}(\Omega)\frac{\partial^2q_{lm}(t)}{\partial 
t^2}=-\sum\limits_{l=0}^{\infty}\sum_{m=-l}^l \frac{\omega_p^2l}{2l+1}Y_{lm}(\Omega)q_{lm}(t)\\
+\omega_p^2\sum\limits_{l=1}^{\infty}\sum\limits_{m=-l}^l\sum\limits_i \frac{l+1}{2l+1}Y_{lm}(\Omega)A_{lm}\int\limits_{0}^a\frac{r_1^{l+2}dr_1}{a^{l+2}}j_l(k_{li}r_1)sin(\omega_{li}t).\\
\end{array}
\end{equation}
From the above equation we notice that for $l=0$ we get $\frac{\partial^2q_{00}}{\partial t^2}=0$ and thus $q_{00}(t)=0$ (as $q(0)=0$ and $\lim_{t\rightarrow \infty} q(t)<\infty$). For $l\geq 1$ we get 
\begin{equation}
    \frac{\partial^2 q_{lm}(t)}{\partial t^2}=-\frac{\omega_p^2 l}{2l+1}q_{lm}(t)+\sum\limits_i\omega_p^2\frac{l+1}{2l+1}
A_{lm} n\int\limits_0^a\frac{r_1^{l+2}dr_1}{a^{l+2}}j_l(k_{li}r_1)sin(\omega_{li}t),
\end{equation}
which requires the solution form,
\begin{equation}
\label{solution}
\begin{array}{l}
    q_{lm}(t)=B_{lm}/a^2 sin(\omega_p\sqrt{\frac{l}{2l+1}}t)\\
+\sum\limits_i A_{lm}\frac{(l+1)\omega_p^2}{\omega_p^2-(2l+1)\omega_{li}^2}n
 \int\limits_0^a\frac{r_1^{l+2}dr_1}{a^{l+2}}j_l(k_{li}r_1)sin(\omega_{li}t),\\
\end{array}
\end{equation}
and $\delta\rho_2(\mathbf{r},t)=\sum\limits_{l=1}^{\infty}\sum\limits_{m=-l}^lq_{lm}(t)Y_{lm}(\Omega)
\delta(a-r)$.
The first term in Eq. (\ref{solution}) describes the self-oscillations of surface plasmons, whereas the second one displays the surface plasmon oscillations induced by the volume plasmons. This induced part of surface oscillations is nonzero  only when the volume modes are excited and their amplitudes, $A_{lmi}$, are nonzero. The frequencies of self-oscillations of the surface plasmons are equal to $\omega_{l0}=\omega_p\sqrt{\frac{l}{2l+1}}$, corresponding to various  multipole modes (numbered with $l$). Note that these frequencies are lower that the bulk plasmon frequency ($\omega_p=\sqrt{\frac{n q^2 4\pi}{m}}$ in Gauss units or $ \sqrt{\frac{n q^2}{\varepsilon_0 m}}$ in SI), whereas the volume plasmon modes oscillate with frequencies higher than $\omega_p$. Worth noting is also an absence of the temperature dependence of the surface ionic plasmon resonances in contrary to the volume plasmon self-frequencies.


\begin{thebibliography}{10}

\bibitem{wzr3a}
K.~Okamoto, I.~Niki, A.~Scherer, Y.~Narukawa, and Y.~Kawakami, ``Surface
  plasmon enhanced spontaneous emission rate of {InGaN/ GaN} quantum wells
  probed by time-resolved photoluminescence spectroscopy,'' {\em Appl. Phys.
  Lett.}~{\bf 87}, p.~071102, 2005.

\bibitem{wzmocn1}
S.~Pillai, K.~R. Catchpole, T.~Trupke, G.~Zhang, J.~Zhao, and G.~M.A,
  ``Enhanced emission from {S}i-based light-emitting diodes using surface
  plasmons,'' {\em Appl. Phys. Lett.}~{\bf 88}, p.~161102, 2006.

\bibitem{wzr2}
D.~M. Schaadt, B.~Feng, and E.~T. Yu, ``Enhanced semiconductor optical
  absorption via surface plasmon excitation in metal nanoparticles,'' {\em
  Appl. Phys. Lett.}~{\bf 86}, p.~063106, 2005.

\bibitem{konk}
S.~P. Sundararajan, N.~K. Grandy, N.~Mirin, and N.~J. Halas,
  ``Nanoparticle-induced enhancement and suppression of photocurrent in a
  silicon photodiode,'' {\em Nano Lett.}~{\bf 8}, p.~624, 2008.

\bibitem{wzmocn2}
M.~Westphalen, U.~Kreibig, J.~Rostalski, H.~L{\"u}th, and D.~Meissner, ``Metal
  cluster enhanced organic solar cells,'' {\em Sol. Energy Mater. Sol.
  Cells}~{\bf 61}, p.~97, 2000.

\bibitem{mof}
A.~J. Morfa, K.~L. Rowlen, T.~H. Reilly, M.~J. Romero, and J.~Lagemaat,
  ``Plasmon-enhanced solar energy conversion in organic bulk heterojunction
  photovoltaics,'' {\em Appl. Phys. Lett.}~{\bf 92}, p.~013504, 2008.

\bibitem{jacak11}
W.~Jacak, J.~Krasnyj, J.~Jacak, R.~Gonczarek, A.~Chepok, L.~Jacak, D.~Hu, and
  D.~Schaadt, ``Radius dependent shift in surface plasmon frequency in large
  metallic nanospheres: {T}heory and experiment,'' {\em J. Appl. Phys.}~{\bf
  107}, p.~124317, 2010.

\bibitem{atwater1}
S.~A. Maier, P.~G. Kik, and H.~A. Atwater, ``Optical pulse propagation in metal
  nanoparticle chain waveguides,'' {\em Phys. Rev. B}~{\bf 67}, p.~205402,
  2003.

\bibitem{vidal2010}
P.~A. Huidobro, M.~L. Nesterov, L.~Martin-Moreno, and F.~J. Garcia-Vidal,
  ``Transformation optics for plasmonics,'' {\em Nano Lett.}~{\bf 10},
  pp.~1985--1990, 2010.

\bibitem{zastos}
S.~A. Maier, {\em Plasmonics: {F}undamentals and {A}pplications}, Springer,
  Berlin, 2007.

\bibitem{maradudin}
A.~V. Zayats, I.~I. Smolyaninov, and A.~A. Maradudin, ``Nano-optics of surface
  plasmon polaritons,'' {\em Phys. Rep.}~{\bf 408}, p.~131, 2005.

\bibitem{deabajo}
F.~J.~G. de~Abajo, ``Optical excitations in electron microscopy,'' {\em Rev.
  Mod. Phys.}~{\bf 82}, p.~209, 2010.

\bibitem{plasmons}
W.~L. Barnes, A.~Dereux, and T.~W. Ebbesen, ``Surface plasmon subwavelength
  optics,'' {\em Nature}~{\bf 424}, p.~824, 2003.

\bibitem{pitarke2006}
J.~M. Pitarke, V.~M. Silkin, E.~V. Chulkov, and P.~M. Echenique, ``Theory of
  surface plasmons and surface-plasmon polaritons,'' {\em Reports on Progress
  in Physics}~{\bf 70}, pp.~1--87, 2007.

\bibitem{berini2009}
P.~Berini, ``Long-range surface plasmon polaritons,'' {\em Advances in Optics
  and Photonics}~{\bf 1}, pp.~484--588, 2009.

\bibitem{lan}
L.~D. Landau and E.~M. Lifshitz, {\em Field {T}heory}, Nauka, Moscow, 1973.

\bibitem{jackson1998}
J.~D. Jackson, {\em Classical Electrodynamics}, John Willey and Sons Inc., New
  York, 1998.

\bibitem{jacak2014b}
W.~Jacak, ``Propagation of collective surface plasmons in
  1{D} periodic ionic structure,''
\newblock submitted.

\bibitem{jacak5}
J.~Jacak, J.~Krasnyj, W.~Jacak, R.~Gonczarek, A.~Chepok, and L.~Jacak,
  ``Surface and volume plasmons in metallic nanospheres in semiclassical
  {RPA}-type approach; near-field coupling of surface plasmons with
  semiconductor substrate,'' {\em Phys. Rev. B}~{\bf 82}, p.~035418, 2010.

\bibitem{rpa}
D.~Pines, {\em Elementary {E}xcitations in {S}olids}, ABP Perseus Books,
  Massachusetts, 1999.

\bibitem{pines}
D.~Bohm and D.~Pines, ``A collective description of electron interactions:
  {III}. coulomb interactions in a degenerate electron gas,'' {\em Phys.
  Rev.}~{\bf 92}, p.~609, 1953.

\bibitem{brack}
M.~Brack, ``The physics of simple metal clusters: self-consistent jellium model
  and semiclassical approaches,'' {\em Rev. of Mod. Phys.}~{\bf 65}, p.~667,
  1993.

\bibitem{ekardt}
W.~Ekardt, ``Size-dependent photoabsorption and photoemission of small metal
  particles,'' {\em Phys. Rev. B}~{\bf 31}, p.~6360, 1985.

\bibitem{mie}
G.~Mie, ``Beitrige zur {O}ptik tr{\"u}ber {M}edien, speziell kolloidaler
  {M}etall{\"o}sungen,'' {\em Ann. Phys.}~{\bf 25}, p.~376, 1908.

\bibitem{atwater}
M.~L. Brongersma, J.~W. Hartman, and H.~A. Atwater, ``Electromagnetic energy
  transfer and switching in nanoparticle chain arrays below the diffraction
  limit,'' {\em Phys. Rev. B}~{\bf 62}, p.~R16356, 2000.

\bibitem{gradst}
I.~S. Gradshteyn and I.~M. Ryzhik, {\em Table of {I}ntegrals {S}eries and
  {P}roducts}, Academic Press, Inc., Boston, 1994.

\end{thebibliography}
\end{document}